\documentstyle[preprint,aps,pre]{revtex}

\begin{document}

\title{``Fuzzy" stochastic resonance: robustness against noise
tuning due to non Gaussian noises}

\author{M. A. Fuentes $^1$\thanks{Electronic address:
fuentesm@cab.cnea.gov.ar}, C. Tessone $^2$ \thanks{Electronic
address: tessonec@venus.fisica.unlp.edu.ar}, H. S. Wio $^1$
\thanks{Electronic address: wio@cab.cnea.gov.ar.} and R. Toral
$^3$\thanks{Electronic address: raul@imedea.uib.es} }
\address{$^1$ Grupo de F\'{\i }sica Estad\'{\i }stica
\thanks{http://www.cab.cnea.gov.ar/Cab/invbasica/FisEstad/estadis.htm},
Centro At\'omico Bariloche \thanks{Comisi\'on Nacional de la
Energ\'{\i}a At\'omica} and Instituto Balseiro \thanks{Comisi\'on
Nacional de la Energ\'{\i}a At\'omica and Universidad Nacional de
Cuyo}, 8400--San Carlos de Bariloche, Argentina, \\ $^2$Instituto
de F\'{\i}sica de La Plata, C.C.
63, Universidad Nacional de La Plata, 1900 La Plata. Argentina. \\
$^3$Departament de F{\'\i}sica, Universitat de les Illes Balears
and IMEDEA (CSIC-UIB)\thanks{http://www.imedea.uib.es}, Palma de
Mallorca, Spain}

\maketitle

\begin{abstract}
We have analyzed the phenomenon of stochastic resonance in a
system driven by {\bf non Gaussian} noises. We have considered
both white and colored noises. In the latter case we have
obtained a consistent Markovian approximation that enables us to
get quasi-analytical results for the signal-to-noise ratio. As
the system departs from Gaussian behavior, our main findings are:
an enhancement of the response together with a marked robustness
against noise tuning. These remarkable findings are supported by
extensive numerical simulations. We also discuss the relation
with some experiments in sensory systems.
\end{abstract}
\vspace{1.0cm}

{\it Stochastic resonance} (SR) has attracted considerable
interest due, among other aspects, to its potential technological
applications for optimizing the response to weak external signals
in nonlinear dynamical systems, as well as to its connection with
some biological mechanisms. There is a wealth of papers,
conference proceedings and reviews on this subject, Ref.
\cite{srrmp} being the most recent and complete one, showing
the large number of applications in science and technology,
ranging from paleoclimatology \cite{BEN,nico}, to electronic
circuits \cite{elect}, lasers \cite{laser}, chemical systems
\cite{chem}, and the connection with some situations of
biological interest (noise-induced information flow in sensory
neurons in living systems, influence in ion-channel gating, in
visual perception or in the human blood pressure regulatory
system) \cite{biol}. Recent works have shown a tendency,
determined by the possible technological applications, pointing
towards achieving an enhancement of the system response (by means
of the coupling of several SR units in what conforms an {\it
extended medium} \cite{buls1,nosot,bowi}), or analyzing the
possibility of making the system response less dependent on a
fine tuning of the noise intensity \cite{claudio}, as well as
different ways to control the phenomenon \cite{control}.

A majority of studies on SR have been made on bistable
one-dimensional double-well systems, and in almost all cases,
with very few exceptions \cite{nG}, the noises are assumed to be
Gaussian. However, some experimental results in sensory systems,
particularly for one kind of crayfish \cite{circle} as well as
recent results for rat skin \cite{nuevo}, offer strong
indications that the noise source in these systems could be non
Gaussian.

In this letter we analyze the case of SR when the noise source is
non Gaussian. We start sketching the study of a particular class
of Langevin equations having non Gaussian stationary distribution
functions \cite{borland1}, a work based on the generalized
thermostatistics proposed by Tsallis \cite{TS} which has been
successfully applied to a wide variety of physical systems
\cite{TS2}. We consider the following problem
\begin{eqnarray}
\dot{x} &=&f(x,t)+g(x)\eta (t)  \label{equis} \\
\dot{\eta} &=&-\frac{1}{\tau }\frac{d}{d{\eta }}V_{q}(\eta )
+\frac{1}{\tau }\xi (t)  \label{nu}
\end{eqnarray}
where $\xi (t)$ is a Gaussian white noise of zero mean and
correlation $\langle\xi (t)\xi (t^{\prime })\rangle= 2\,D \delta
(t-t^{\prime })$, and $V_{q}(\eta )$ is given by \cite{borland1}
\begin{equation}
V_{q}(\eta )=\frac{1}{\beta (q-1)}\ln [1+\beta (q-1)\frac{\eta ^{2}}{2}],
\end{equation}
where $\beta =\frac{\tau}{D}$. The function $f(x,t)$ is derived
from a potential $U(x,t)$, consisting of a double well potential
and a linear term modulated by  $S(t)= F\,\cos (\omega t)$
($f(x,t)=-\frac{\partial U}{\partial x}=-U_0^{\prime} + S(t)$).
For $F=0$ this problem corresponds to the case of diffusion in a
potential $U_0(x)$, induced by  $\eta$, a colored non-Gaussian
noise. The stationary probability distribution for the random
variable $\eta$ is given by $P_{q}^{st}(\eta)=Z_{q}^{-1}\left[
1+\beta (q-1)\frac{\eta ^{2}}{2}\right] ^{\frac{-1}{q-1}}$ where
$Z_{q}$ is the normalization factor. Clearly, when $q \to 1$ we
recover the limit of $\eta $ being a Gaussian colored noise
(Ornstein-Uhlenbeck process). For $q<1$ the above distribution
cuts-off the values of the $\eta$ process to
$\mid\eta\mid<[(1-q)\frac{\beta}{2}]^{-1/2}$. It is easy to show
that the stationary mean value and variance are given,
respectively, by $\langle \eta\rangle=0$ and $\langle \eta^2
\rangle=[\beta(5-3q)]^{-1}$, the latter diverging for all $q\ge
5/3$.

Applying the path-integral formalism to the Langevin equations
given by Eqs. (\ref{equis}-\ref{nu}), and making an
adiabatic-like elimination procedure \cite{pi1,pi2,pi3} it is
possible to arrive at an {\it effective Markovian approximation}.
The specific details are shown elsewhere \cite{nogaus}. Such an
approximation yields the following Fokker--Planck equation (FPE)
for the evolution of the probability $P(x,t)$
\begin{equation}
\partial _{t}P(x,t)=-\partial _{x}[A(x)P(x,t)]+\frac{1}{2}\partial
_{x}^{2}[B(x)P(x,t)].  \label{FPaprox}
\end{equation}
Where
\begin{equation}
A(x)=\frac{U_0^{\prime }} {\Bigl( \frac{1-\frac{\tau }{2 D}
(q-1)U_0^{\prime }{}^{2}} {1+\frac{\tau }{2 D}(q-1)U_0^{\prime
}{}^{2}} \Bigr) + \tau U_0^{\prime \prime }[1+\frac{\tau }{ 2
D}(q-1)U_0^{\prime }{}^{2}]}
\end{equation}
and
\begin{equation}
B(x)=D \, \Biggl( \frac{[1+\frac{\tau }{2 D}(q-1)U_0^{\prime
}{}^{2}]^2}{\tau U_0^{\prime \prime }[1+\frac{\tau }{2
D}(q-1)U_0^{\prime }{}^{2}]^2+ [1-\frac{\tau }{2
D}(q-1)U_0^{\prime }{}^{2}] }\Biggr) ^{2}.
\end{equation}
The stationary distribution of the FPE in Eq. (\ref{FPaprox}) is
thus
\begin{equation}
P^{st}(x)=\frac{{\cal N}}{B(x)}\exp \left[ -\Phi (x)\right]
\label{estacionaria}
\end{equation}
where ${\cal N}$ is the normalization factor, and
\begin{equation}
\Phi (x)=2 \, \int^x \frac{A(x')}{B(x')}dx'.
\end{equation}
The indicated FPE and its associated stationary distribution
enable us to obtain the mean-first-passage-time (MFPT) through a
Kramers like approximation \cite{HG}. This quantity is the
necessary ingredient to work within {\it the two-state model}
(TST) \cite{nico,McNam}. The MFPT to reach $x=x_0$ starting from
$x=a$ can be obtained from
\begin{equation}
\label{kla}
T(x_{0})=\int_{a}^{x_{0}}dy\,{\rm e}^{\Phi(y)}\int_{-\infty }^{y}dz\,B(z)^{-1}{\rm e}^{-\Phi(z)},
\end{equation}
We focus on polynomial-like forms for the potential, adopting
\begin{equation}
U(x)=\frac{x^{4}}{4}-\frac{x^{2}}{2}+S(t) x,  \label{cuartico}
\end{equation}
setting $S(t)= F\, \cos(\omega t),$ and assuming that $\omega
^{-1}$ is large compared to the characteristic relaxation times
in both wells.

We follow the standard TST approach \cite{McNam} in order to
obtain $R$, the signal to noise ratio (SNR). In our case the
result for $R$, defined as the ratio of the strength of the
output signal to the broadband noise output evaluated at the
signal frequency, is
\begin{equation}
\label{R} R=\left .  \frac{F^2
\pi}{2\,T^3}\left[\frac{dT}{dS(t)}\right]^2 \right|_{S(t)=0}.
\end{equation}
with $S(t)$ the applied signal.

We present the results for the SNR obtained evaluating Eq.
(\ref{R}). In Fig. 1a we depict the SNR vs. the noise intensity
$D_c$, for a fixed value of the time correlation $\tau$ and
various $q$, while Fig. 1b shows the equivalent plot for a fixed
value of $q$ and several values of $\tau $. However, $D_c$ the
noise intensity of the non Gaussian process is related to the
Gaussian white noise $D$ through the scaling $D_C= D
[5-3q]^{-1}$, that was the scaling we used. In the former case
the general trend is that the maximum of the SNR curve increases
when $q < 1$, i.e. when the system departs from the Gaussian
behavior. In the latter case the general trend agrees with the
results for colored Gaussian noises \cite{color}, where it was
shown that the increase of the correlation time induces a
decrease of the maximum of SNR as well as its shift towards
larger values of the noise intensity. The latter fact is a
consequence of the suppression of the switching rate with
increasing $\tau$. Both qualitative trends are confirmed by
extensive numerical simulations of the system in Eqs.
(\ref{equis}-\ref{nu}), which has been numerically integrated
using a stochastic Runge-Kutta-type method \cite{raul}. The
results were obtained averaging over $2000$ trajectories ($5000$
trajectories for $\tau = 0$). Fig. 2a shows the simulation
results for the same situation and parameters indicated in Fig.
1a. Here, in addition to the increase of the maximum of the SNR
curve for values of $q < 1$, we also found another remarkable
aspect that is not reproduced or predicted by our effective
Markovian approximation. It is the fact that the maximum of the
SNR curve flattens for lower values of $q$, indicating that the
system, when departing from Gaussian behavior, requires a less
fine tuning of the noise intensity in order to maximize its
response to a weak external signal. Fig. 2b shows the simulation
results for the same situation and parameters indicated in Fig.
1b. Again we found an agreement with the general behavior found
for colored Gaussian noises \cite{color}.

In order to compare with some experiments where the need to use
non-Gaussian noises has been suggested, we consider the results
obtained for a crayfish in Ref.\cite{circle}. A qualitative
comparison of our results in Figs. 1 and 2 with those given there
shows that the agreement between theory and experiment does indeed
improve by the use of non-Gaussian noises corresponding to $q<1$.
However, it is known that rather than applying our simple
one-variable model, to the crayfish neural system under
consideration it is better to use the celebrated FitzHugh-Nagumo
one. In the same way as in Ref.\cite{circle}, we define the model
system through the set of equations
\begin{eqnarray}
\tau_v \dot{v} &=& v (v- 0.5) (1-v) - w, \nonumber \\
\tau_w \dot{w} &=& v - w + \epsilon \cos(\Omega t) + \eta(t),
\label{fhn-eq}
\end{eqnarray}
where $v(t)$ is the variable associated to the action potential
(in an activator-inhibitor model, it corresponds to the activator
variable), and $w(t)$ is the recovery (inhibitor) variable. The
characteristic times for such variables are, respectively,
$\tau_v$ and $\tau_w$. The recovery variable is submitted to a
periodic signal (whose (small) amplitude we denote by $\epsilon$
and its frequency by $\Omega$) and a noise source $\eta(t)$. This
random variable, now chosen as white (that is with $\tau = 0$), is
distributed accordingly to the stationary pdf of Eq. (\ref{nu}).
We have integrated numerically Eqs. (\ref{fhn-eq}) using a
stochastic Runge-Kutta-type method with a time step of $\Delta t=1
\hbox{ms}$.

The time series $v(t)$ was converted into a spike train
(mimicking what happens in the nervous system). One spike occurs
when $v(t)$ exceeds the threshold potential $b_c$. Each spike was
modeled as a square wave of height $V_s=1 \hbox{V}$, and
duration $t_s=3 \hbox{ms}$. It is not possible to have two
successive spikes with a lag smaller than $t_s$. The SNR was
evaluated following the usual procedure on this modified signal.

Along all the simulations, we fixed $\tau_v=10^{-5} \hbox{s}$
and $\tau_w=10^{-2} \hbox{s}$. While we considered the signal as
having frequency $\Omega=55 \hbox{Hz}$ and amplitude $\epsilon =
0.03 \hbox{V}$. All these values agree with those from Ref.
\cite{circle}, however  their results are a particular case of
ours, i.e. $q=1$, for the noise source.

We have compared our simulations for $R$, the SNR, as a function
of noise standard deviation for different values of $q$ and
$b_c$, the threshold potential, with the experimental data in
\cite{circle}. Here again we have scaled the noise intensity
according to $D_n=D [5-3q]^{-1}$, where $D_n$ is the intensity of
the Non Gaussian noise while $D$corresponds to the pdf parameter.
Figure 3 shows the results obtained for $b_c=0.18 \hbox{V}$. We
see that the (Gaussian) case $q=1$ underestimates the measured
SNR values for the whole range of noise intensities, while
$q=0.75$ exhibits an overestimation for large noise values.
However, if we adopt $q=0.47$ the system exhibits a nice
agreement with experimental data both near the SNR peak and for
large values of noise. When the threshold value is fixed at
$b_c=0.15\hbox{V}$ as in \cite{circle} (results not shown here),
the case $q=1$ corresponds exactly to the theoretical fit in
\cite{circle}. From the comparison with $q \neq 1$ it becomes
apparent that, although the fit is acceptable near the $R$
maximum, for large noise intensities the $q=1$ case underestimate
the values of $R$. By setting $q=0.75$, we find a much better fit
even in the region of large noises, and when $q=0.30$ and for
large values of the noise standard deviation, the results exhibit
values above the measured ones.

It it worth remarking here that the experimental behavior in the
low noise limit is not reproduced by any model. This is due to the
spontaneous firing of neurons in the real crayfish neural system
\cite{circle}.

Summarizing, motivated by some experimental results in sensory
systems \cite{circle,nuevo}, we have analyzed the problem of SR
when the noise source is non Gaussian. We have chosen a non
Gaussian noise source (white or colored) with a probability
distribution based on the generalized thermostatistics \cite{TS}.
In the colored case and making use of a path integral approach,
we have obtained an effective Markovian approximation that allows
us to get some analytical results. In addition, we have performed
exhaustive numerical simulations. Even though the agreement
between theory and numerical simulations is only partial and
qualitative, the effective Markovian approximation turns out to
be extremely useful to predict (qualitatively) general trends in
the behavior of the system under study.

Our numerical and theoretical results indicate that: (i) for a
fixed value of $\tau $, the maximum value of the SNR increases
with decreasing $q$; (ii) for a given value of $q$, the optimal
noise intensity (that one maximizing SNR) decreases with $q$ and
its value is approximately independent of $\tau $; (iii) for a
fixed value of the noise intensity, the optimal value of $q$ is
independent of $\tau $ and in general it turns out that $q \neq
1$.

As we depart from Gaussian behavior (with $q<1$), the SNR shows
two main aspects: firstly its maximum as a function of the noise
intensity increases, secondly it becomes less dependent on the
precise value of the noise intensity. Both aspects are of great
relevance for technological applications \cite{srrmp}. However,
as was indicated in Ref. \cite{nuevo}, non Gaussian noises could
be an intrinsic characteristic in biological systems,
particularly in sensory systems \cite{biol,circle,nuevo}. In
addition to the increase in the response (SNR), the reduction in
the need of {\it tuning} a precise value of the noise intensity
is of particular relevance both in technology and in order to
understand how a biological system can exploit this phenomenon.
As an example of this, simulations of a FHN model with non
Gaussian noise shows a behavior in better agreement with the
experimental results in a crayfish sensory system than simulations
made with a Gaussian noise source. The present results indicate
that the noise model used here offers an adequate framework to
analyze such a problem.

\noindent {\bf Acknowledgments:} The authors thank C. Tsallis for
useful discussions and V. Grunfeld for a revision of the
manuscript. HSW acknowledges support from CONICET, Argentine
agency. RT acknowledges financial support from DGES, project
numbers BFM2000-1108 and PB97-0141-C02-01.

\vspace{2.0cm}

\begin{description}
\item[Figure 1:]  Theoretical value of the SNR, $R$, vs. the noise
intensity $D$, for: (a) $\tau = 0.1$ and the following values of
$q =0.25, 0.75, 1.0, 1.25$ (continuous line, from top to bottom).
(b) $q=0.75$ and the following values of $\tau =0.25, 0.75, 1.5$
(broken line from top to bottom). We adopted $F=0.1$ and $\omega=
0.1$.

\item[Figure 2:] Simulation results of the SNR, $R$, vs. the
noise intensity $D$, for (a) $\tau = 0.1$ and the following
values of $q =0.25, 0.75, 1.0, 1.25$ (from top to bottom). (b)
$q=0.75$ and the following values of $\tau =0.25, 0.75, 1.5$
(from top to bottom). We adopted $F=0.1$ and $\omega= 0.1$.

\item[Figure 3:] The SNR, $R$, for the FHN model is plotted
as a function of $\sqrt{2 D}$ (the noise standard deviation). The
different curves show the results for different $q-$values: $q=1$
($\bullet$), $q=0.75$ ($\diamondsuit $), $q=0.47$ ($\nabla $). In
all the cases we fixed $b_c = 0.18 \hbox{V}$, simulation
results are compared with the experimental data ($\Box $).
\end{description}


\end{document}